\documentclass{desyproc}
\begin{document}
\newcommand{\M}{\mbox{m}}
\newcommand{\n}{\mbox{$n_f$}}
\newcommand{\EP}{\mbox{$e^+$}}
\newcommand{\EM}{\mbox{$e^-$}}
\newcommand{\EPEM}{\mbox{$e^+e^{-}$}}
\newcommand{\EMEM}{\mbox{$e^-e^-$}}
\newcommand{\GG}{\mbox{$\gamma\gamma$}}
\newcommand{\GE}{\mbox{$\gamma e$}}
\newcommand{\GP}{\mbox{$\gamma e^+$}}
\newcommand{\TEV}{\mbox{TeV}}
\newcommand{\GEV}{\mbox{GeV}}
\newcommand{\LGG}{\mbox{$L_{\gamma\gamma}$}}
\newcommand{\LGE}{\mbox{$L_{\gamma e}$}}
\newcommand{\LEE}{\mbox{$L_{ee}$}}
\newcommand{\LEPEM}{\mbox{$L_{e^+e^-}$}}
\newcommand{\WGG}{\mbox{$W_{\gamma\gamma}$}}
\newcommand{\WGE}{\mbox{$W_{\gamma e}$}}
\newcommand{\EV}{\mbox{eV}}
\newcommand{\CM}{\mbox{cm}}
\newcommand{\MM}{\mbox{mm}}
\newcommand{\NM}{\mbox{nm}}
\newcommand{\MKM}{\mbox{$\mu$m}}
\newcommand{\SEC}{\mbox{s}}
\newcommand{\CMS}{\mbox{cm$^{-2}$s$^{-1}$}}
\newcommand{\MRAD}{\mbox{mrad}}
\newcommand{\IND}{\hspace*{\parindent}}
\newcommand{\E}{\mbox{$\epsilon$}}
\newcommand{\EN}{\mbox{$\epsilon_n$}}
\newcommand{\EI}{\mbox{$\epsilon_i$}}
\newcommand{\ENI}{\mbox{$\epsilon_{ni}$}}
\newcommand{\ENX}{\mbox{$\epsilon_{nx}$}}
\newcommand{\ENY}{\mbox{$\epsilon_{ny}$}}
\newcommand{\EX}{\mbox{$\epsilon_x$}}
\newcommand{\EY}{\mbox{$\epsilon_y$}}
\newcommand{\BI}{\mbox{$\beta_i$}}
\newcommand{\BX}{\mbox{$\beta_x$}}
\newcommand{\BY}{\mbox{$\beta_y$}}
\newcommand{\SX}{\mbox{$\sigma_x$}}
\newcommand{\SY}{\mbox{$\sigma_y$}}
\newcommand{\SZ}{\mbox{$\sigma_z$}}
\newcommand{\SI}{\mbox{$\sigma_i$}}
\newcommand{\SIP}{\mbox{$\sigma_i^{\prime}$}}
\newcommand{\be}{\begin{equation}}
\newcommand{\ee}{\end{equation}}
\newcommand{\bc}{\begin{center}}
\newcommand{\ec}{\end{center}}
\newcommand{\bi}{\begin{itemize}}
\newcommand{\ei}{\end{itemize}}
\newcommand{\ben}{\begin{enumerate}}
\newcommand{\een}{\end{enumerate}}
\newcommand{\bm}{\boldmath}

%------------------------------------
\title{Photon Collider Technology Overview}

%for single authors the superscripts are optional
\author{{\slshape V.~I.~Telnov}\\[1ex]
Institute of Nuclear Physics, Novosibirsk 630090, Russia}

% please enter yoru conference contribution ID from the INDICO webpage below
\contribID{13}

% please do not modify the following four lines
\confID{1407}  % if the conference is on Indico uncomment this line
\desyproc{DESY-PROC-2009-03}
\acronym{PHOTON09} % if you want the Acronym in the page footer uncomment this line
\doi  % if there is an online version we will register DOIs

\maketitle

\begin{abstract}
  In this conference paper, I review the present status and technical problems of the Photon collider, as well as various additional applications of Compton scattering.

\end{abstract}

\section{Intoduction}

In this report, I provide an overview of the technical (and some political) aspects of the Photon Linear Collider (PLC). The physics program at the PLC is discussed in K.~M\"{o}nig's talk at this conference~\cite{Monig-photon09}.

The photon collider based on the conversion of electrons at a linear collider to high-energy photons through Compton scattering of laser photons has been discussed and developed since early 1980s~\cite{GKST81,GKST83-84}. A photon collider would be a very natural and relatively cheap supplement to a high-energy \EPEM\ linear collider. It would allow the study of New Physics in two additional types of collisions, \GG\ and \GE\, with energies and luminosities close to those in \EPEM\ collisions.  A comprehensive description of the PLC is given in the TESLA TDR \cite{TESLATDR}; practically everything regarding the photon collider at TESLA is valid for the PLC at the ILC. Further progress on the PLC after 2001 has been summarized in my talks at PHOTON2005~\cite{TELacta1,TELacta2} and PHOTON2007~\cite{Tel-Photon07}.

What's new since 2007?  Unfortunately, the future of the ILC is still highly uncertain. A lot depends on the physics results from the LHC, but even the discovery of a new physics at the LHC would not guarantee the approval of the ILC (or CLIC?) construction due to its high cost (``high'' as perceived by politicians). A possible way to overcome this barrier could be to build the linear collider in several stages. Recently, in October 2008, Prof. Hirotaka Sugawara suggested to the ILC Steering Committee the construction of a ``Photon collider Higgs factory as a precursor to ILC,'' as the required energy for producing a 120 GeV Higgs is lower in \GG\ collisions than in \EPEM, positrons are not needed, and therefore such a collider would be much cheaper. While laudable as an attempt to find a way out of the ILC stalemate, this suggestion has caused concern to many in the ILC community because it would have meant an additional delay of 5-6 years in the start of \EPEM\ operations at the ILC. After consultations with PLC experts and additional study of the technical aspects and physics program for a low-energy startup scenario, the ILCSC rightfully concluded that it would more preferable to start with \EPEM\ at $2E=230$ \GEV\ and investigate the Higgs in the $\EPEM\ \to ZH$ process. This option's cost is not much higher than that of a 120 GeV PLC, but the physics case is stronger.

The other interesting activities in the last few years related to the photon
collider and based on ideas originally proposed in the context of
photon colliders are the developments of laser systems for various
applications based on Compton scattering.

The outline of this paper is as follows.  Basic properties of the PLC are considered in Sect.~2.
Technical aspects are overviewed in Sect.~3. The proposal of the PLC as the first stage of the ILC is discussed in Sect.~4.
Various applications of Compton scattering are considered in Sect.~5.

\section{Basics of the photon collider}

Let us consider briefly the main characteristics of backward Compton
scattering important for the photon collider.

\underline{Kinematics}. In the conversion region, a laser photon of energy $\omega_0$
collides with a high-energy electron of energy $E_0$ at a small
collision angle $\alpha_0$ (almost head-on).  The energy of the
scattered photon $\omega$ depends on the photon scattering angle
$\vartheta$ with respect to the initial direction of the electron as
follows~\cite{GKST83-84}:
\begin{equation}
\omega = \frac{\omega_m}{1+(\vartheta/\vartheta_0)^2},\,\,\,\omega_m=\frac{x}{x+1}E_0; \;\;\;\;\vartheta_0= \frac{mc^2}{E_0}
\sqrt{x+1}; \,\,\, x=\frac{4E \omega_0 }{m^2c^4}\cos^2\frac{\alpha_0}{2} \simeq
 19\left[\frac{E_0}{\TEV}\right] \left[\frac{\mu
 \mbox{m}}{\lambda}\right],
\label{kin1}
\end{equation}
where $\omega_m$ is the maximum energy of scattered photons.
For example: $E_0 = 250$ GeV, $\omega_0 = 1.17$ eV ($\lambda=1.06$
\MKM) (for the most powerful solid-state lasers) $\Rightarrow$ $x=4.5$
and $\omega_m/E_0 = 0.82$.  Spectra of scattered photons are broad with enhancement at maximum energies. Formulae for the Compton cross section and graphs can be found elsewhere~\cite{GKST83-84,TESLATDR}.

\underline{Monochromatization.} By collimating the photon beam, one can obtain
monochromatic gamma (or X-ray) beams, which is important for
various potential applications. At the photon collider, there are no collimators,
but there is some monochromatization of collisions due to the fact that the
higher-energy photons collide at smaller spot sizes and contribute
more to the luminosity.  In \GE\ collisions, the resulting
luminosity spectrum can, in principle, be very narrow (the
electron beam collides with the most high-energy photons), while in
\GG\ collisions the resulting luminosity spectra have the width at
half-maximum of about 10-15\% ~\cite{TEL90,TEL95}.

\underline{Maximum energy of scattered photons}. With increasing $x$,
the energy of the backscattered photons increases and the energy
spectrum becomes narrower.  However, at large values of $x$ photons
may be lost due to creation of \EPEM\ pairs in collisions with laser
photons, which leads to a reduction of the \GG\ luminosity~\cite{GKST83-84,TEL90,TEL95}.  The threshold of this reaction
is $x=2(1+\sqrt{2})\approx 4.83$.  The corresponding wavelength of laser photons is $\lambda= 4.2 E_0 \;[\TEV]\;\; \MKM\,$.
Hence, the maximum energy of photons at the PLC is about $0.8E_0$.

\underline{Polarization} If laser photons are 100\% circularly polarized, the
backscattered photons at the highest photon energy also have
100\% circular polarization (even for unpolarized electrons and for
any value of $x$).  The energy spectrum of scattered photons depends on
the average electron helicity $\lambda_{e}$ and that of the laser
photons $P_c$.  The relative number of hard photons increases when one uses beams with a negative value of the product $\lambda_{e} P_c$.
For large $x$, the polarization of electrons increases the number of photons in the high-energy peak almost by a factor of 2 (or 4 in the \GG\ luminosity).
The energy spectrum of the scattered photons for various helicities of the
electron and laser beams can be found elsewhere~\cite{GKST83-84,TESLATDR}.
A high degree of photon's circular polarization is essential
for the study of many physics processes, for example, for suppression
of QED background in the study of the Higgs boson~\cite{TESLATDR}.
The ratio $L_0/L_2$ (0,2 is the total helicity of colliding photons) is larger when electron beams have a higher degree of longitudinal polarization~\cite{TESLATDR}. Modern electron guns give polarization up to $\sim 85$\%, which is OK. A high electron polarization is practically mandatory for the photon collider. With unpolarized electron beams, it would be practically impossible to study the Higgs(120) (or even to observe it).

\underline{Nonlinear effects in the conversion.}
In order to convert nearly all electrons to high-energy photons, the density of laser photons at the conversion point should be so high that the electron can interact with several laser photons simultaneously.  This nonlinear effect is characterized by the parameter $\xi^2 = 2 n_{\gamma} r_e^2 \lambda/\alpha$, where $n_{\gamma}$ is the density of laser photons, $r_e=e^2/mc^2$ and $\alpha=e^2/\hbar c$~(\cite{TESLATDR} and references therein). The transverse motion of the electron in the electromagnetic wave leads to an effective increase of the electron mass: $m^2 \to m^2(1+\xi^2)$, which decreases the maximum energy of the scattered photons: $\omega_m/E_0 = x/(1+x+\xi^2)$. The Compton spectrum is shifted towards lower energies, higher harmonics appear, and the \GG\ luminosity spectra become broader.  At $x=4.8$, the value of $\omega_m/E_0$ decreases by about 5\% for $\xi^2=0.3$, which can be considered the limit.

\underline{Laser flash energy.}
For small conversion coefficients $k = N_{\gamma}/ N_e \sim 1- \exp(-A/A_0)$, where the flash energy $A_0$ is determined by the diffractive divergence of the laser beam and geometric size of the electron beam. For head-on collisions and very narrow electron beams $A_0 \sim 2 \hbar c \sigma_z/\sigma_c$~\cite{TEL95}, where $\sigma_z$ is the r.m.s.\ length of the electron beam and $\sigma_c$ is the Compton cross section. For $x=4.8$ $A_0 \sim 3 \sigma_z$  [mm] J, while for $x \ll 1$ the Compton cross section approaches the Thomson one and the coefficients $3 \Rightarrow 1$. For the ILC ($\sigma_z=0.3$) mm this estimate gives $A_0 \sim 1$ J.
However, when $k \sim 1$ the nonlinear effects become important. In order to keep $\xi^2$ small one should make the conversion length longer, which increases the required flash energy. In calculating the flash energy, one should also take into account the collision angle between the laser and electron beams (when the laser optics is outside of the electron beam), the effective transverse size of the electron beam due to the tilt in the crab-crossing scheme of collisions and the angular size of the first quad (if optical mirrors are situated outside the detector). A realistic calculation for ILC(500) gives $A_0 \sim 9 J$~\cite{Klemz2005,TEL-Snow2005,TELacta2}.

\section{Technical problems of photon colliders}.
\vspace{-1cm}
\subsection{\GG, \GE\ luminosities}
In \EPEM\ collisions, the maximum achievable luminosity is determined
by beamstrahlung and beam instabilities.  At photon colliders, the
only effect that restricts the \GG\ luminosity is the conversion of
the high-energy photons into \EPEM\ pairs in the field of the opposing
beam -- that is, the coherent pair creation~\cite{ChenTel,TEL90}. For \GE\
collisions, the luminosity is determined by beamstahlung, coherent
pair creation, and the beam displacement during the
collision~\cite{TEL95,TESLATDR}. It is interesting to note that at the center-of-mass energies below
0.5--1 TeV and for electron beams that are not too short (the case of
ILC), coherent pair creation is suppressed due to the broadening and
displacement of the electron beams during the collision.  For \EPEM,
the minimum horizontal beam size restricted by beamstrahlung is about
500 nm at the ILC, while the photon collider can work even with
$\sigma_x \sim 10$ nm at $2E_0=500$ GeV, delivering a luminosity much higher than that
in \EPEM\ collisions~\cite{Tfrei,TEL2001,TESLATDR}.
In fact, the \GG\ luminosity is simply proportional to the {\it geometric} \EMEM\ luminosity $L_{geom}$.

Unfortunately, the beam emittances in the damping-ring designs
currently under consideration cannot achieve beam sizes that are
smaller than $\sigma_x \sim$ 250 nm and $\sigma_y \sim 5$ nm
\cite{TELacta2}, though a reduction of $\sigma_x$ by a factor of two
seems possible.  In principle, one can use electron beams directly
from low-emittance photo-guns, avoiding the need for damping rings
altogether, but at present they offer a product of the transverse
emittances that is noticeably larger than can be obtained with damping
rings: with polarized electron beams directly from photo-guns, the luminosity would be 100 times smaller!

Here is an approximate rule: the luminosity in the
high-energy peak $\LGG \sim 0.1 L_{ \rm
geom}$~\cite{TESLATDR}. With ``nominal'' ILC beam parameters, the expected \GG\ luminosity in the high-energy peak of the luminosity spectrum  $\LGG(z> 0.8z_m) \sim 3.5 \times 10^{33}$ \CMS\ $\sim 0.17\,\LEPEM$ ~\cite{TEL-Snow2005,TELacta2}.

Taking into account the fact that cross sections for many interesting
processes are larger in \GG\ collisions than those in \EPEM\ by an
order of magnitude~\cite{TESLATDR}, the event rate in \GG\ collisions with the nominal ILC beams would be similar, or perhaps somewhat larger, than in \EPEM\ collisions. However, it is a highly unsatisfying situation to have the \GG\
luminosity limited by the beam emittances, an order of magnitude below
its physics limit determined by collision effects.  It is an
extremely interesting and important task to search for a realistic technical
solution for obtaining beams with smaller emittances, and the first order of business should be trying to optimize the damping rings for the specific
requirements of achieving the highest possible luminosity at the
photon collider, as it was emphasized in~\cite{TEL-Snow2005,TELacta2,TEL-LCWS06-2}. Up to now the ILC damping-ring design has been guided only by the baseline \EPEM\ collisions.

\begin{figure}[hbt]
\centering
\vspace*{-0.6 cm}
\hspace{-0.5cm}   \epsfig{file=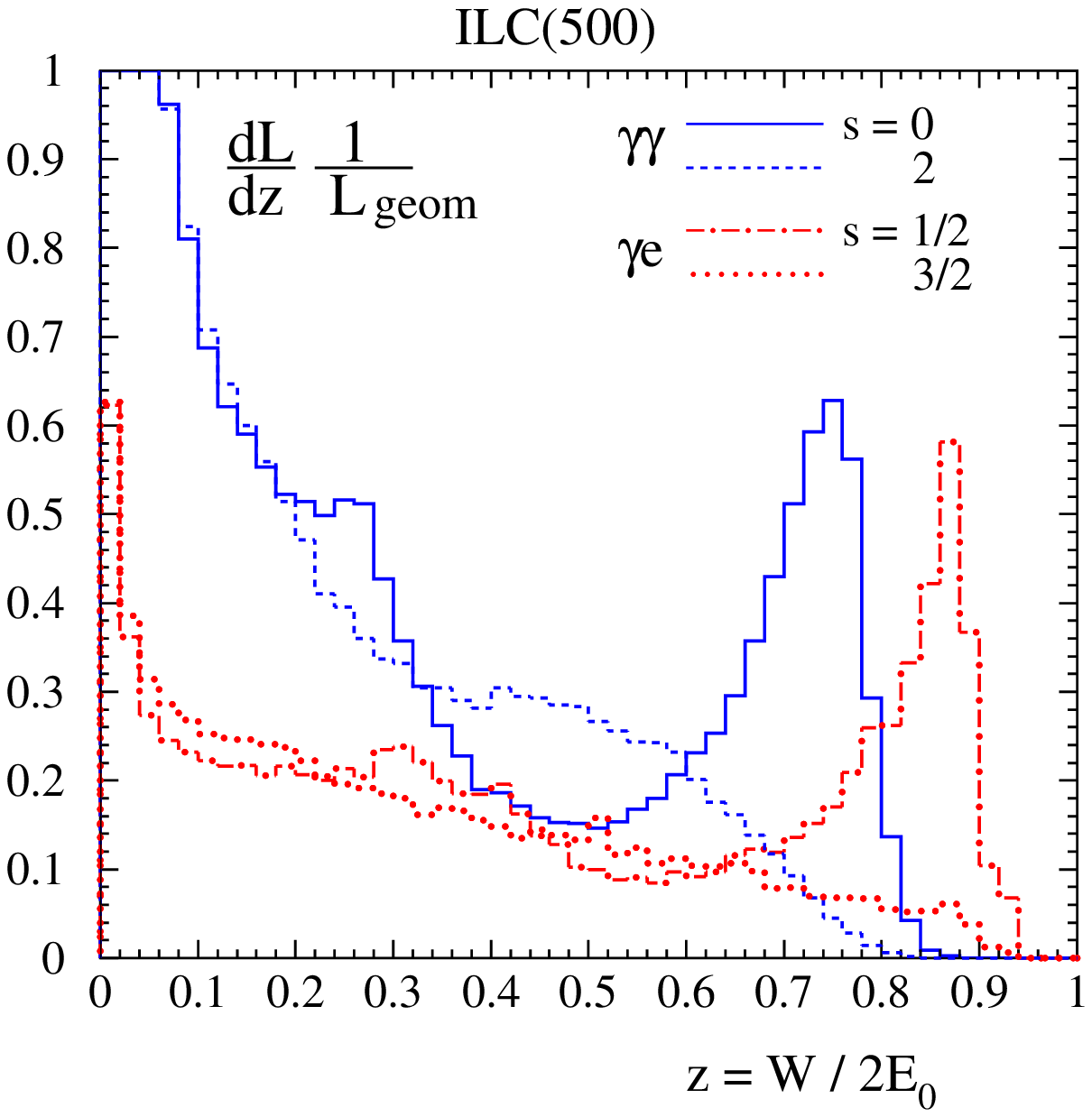,width=7.cm,angle=0}
\hspace{-1cm} \epsfig{file=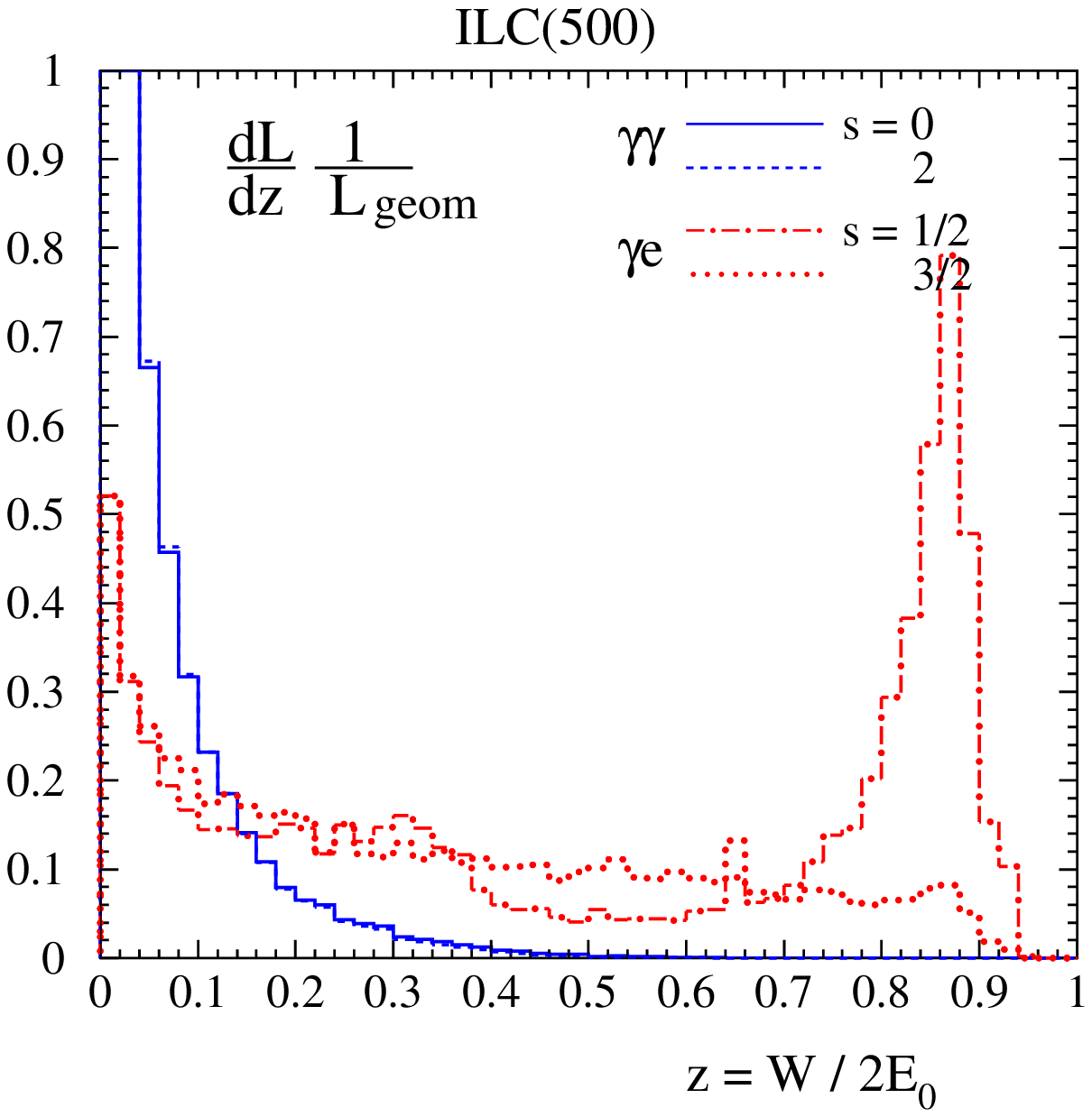,width=7.cm,angle=0}
\vspace{-0.8cm}
\caption{ \GG, \GE\ luminosity spectra, left: both beams are converted
  to photons; right: only one beam is converted to photons.}
 \vspace{-0.3cm}
\label{lumspectra}
\end{figure}

The typical \GG, \GE\ luminosity spectra for the TESLA-ILC(500)
parameters are shown in Fig.~\ref{lumspectra}~\cite{TESLATDR}.
One can see that \GG\ and \GE\
luminosities are comparable and these processes can be studied
simultaneously. However, it is much better to study \GE\ collisions
when only one of the electron beams is converted to photons. In this
case, one can measure the \GE\ luminosity much more
precisely~\cite{Pak}. The problem of measuring the \GE\ luminosity
spectra when both beams are converted to photons  is
due to the uncertainty which direction the photon came from.
For most measurements, the luminosity as high as possible is desired.
However, sometimes very clean \GG\ collisions with good
monochromaticity and a reduced luminosity (in order to avoid
overlapping events) are needed. At large CP--IP distances and a non-zero crossing angle, the detector field serves as a deflecting magnet and allows more-or-less clean and quite monochromatic \GG, \GE\ collisions to be obtained with a reduced luminosity, which will be useful to QCD
studies, \cite{TEL-mont,TEL-ggLHC}.

\underline{Luminosity stabilization.}
Beam collisions (luminosity) at linear colliders can be adjusted by a
feedback system that measures the beam-beam deflection using beam
position monitors and corrects beam positions by fast kickers.  This
method is considered for \EPEM\ collisions and is assumed for \GG\ as
well~\cite{TESLATDR,TELacta2}, though there are some differences
between the \EPEM\ and \GG\ cases.   This problem and a
stabilization algorithm were considered in detail in
Ref.~\cite{TELacta2}.

\underline{Luminosity measurement.}
The measurement of the luminosity at the photon collider is not an
easy task. The spectra are broad, and one should measure the luminosity
and polarization as a function of energies $E_1, E_2$ of the colliding
particles~\cite{Pak}.  The luminosity spectrum and polarization can be
measured using various QED processes. These are  $\GG\to l^+l^-$
($l=e,\mu$)~\cite{TESLATDR,Pak}, $\GG\to
l^+l^-\gamma$~\cite{Pak,Makarenko}  for \GG\ collisions and  $\GE\to\GE$ and
$\GE\to e^-\EPEM$ for \GE\ collisions~\cite{Pak}. Some other SM
processes could be useful as well.

\underline{Absolute beam energy measurement.}
At the photon collider, the edge energy of the photon spectra and the
electron beam energy $E_0$ are not strictly connected due to nonlinear
effects in Compton scattering. The absolute energy calibration of the detector can be done using
the process $\GE\ \to eZ$ (during normal runs in \GE\ mode or mixed \GG\ and \GE\ mode)~\cite{TEL-TILC09-1}.

\subsection{Removal of used beams}

\begin{figure}[!htb]
\vspace{-0.5cm}
%\centering \vspace*{0.3cm}
\begin{minipage}{0.53\linewidth}
\vspace{0.3cm}
The general scheme of the photon collider is shown in
Fig.~\ref{ggcol}. The optimum $b \sim \gamma \sigma_y$,
which is $\sim 1.5$ mm for $\sigma_y=3$ nm and  $2E_0=500$ GeV. This space is too small to fit any kind of a magnet for deflection of used electron beams.  In this case, there is a mixture of \GG, \GE\ and \EMEM\ collisions.
After crossing the conversion region, the electrons have a very broad
energy spectrum, $E=($0.02--1)\,$E_0$ and large disruption
angles due to deflection of low-energy electrons in the field of the opposing beam. The removal of such a beam from the detector is therefore far
from trivial.
\end{minipage}
\begin{minipage}{0.47\linewidth}
\hspace{1cm} \includegraphics[width=5.cm,angle=0]{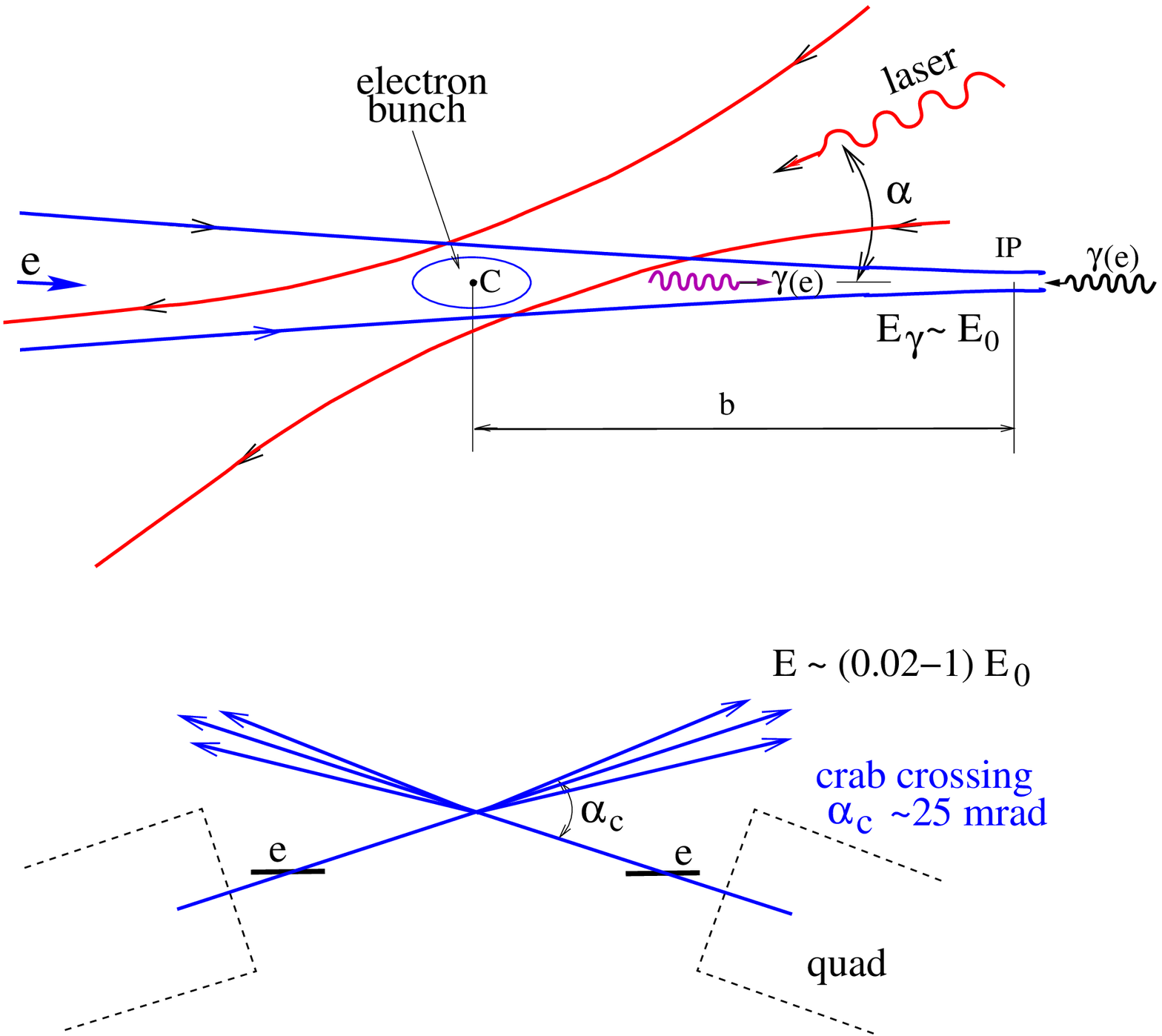}
\caption{Scheme of  \GG, \GE\ collider.}
\label{ggcol}
\end{minipage}
\end{figure}

\vspace{-0.3cm}
The ``crab crossing'' scheme of beam collisions solves the
problem of beam removal at photon colliders~\cite{TEL90,TEL95},
Fig.~\ref{ggcol} (bottom).  In the crab-crossing scheme~\cite{Palmer}, the beams are collided at a crossing angle $\alpha_c$.  In order to preserve the
luminosity, the beams are tilted by a special RF cavity by the angle
$\alpha_c/2$.  If the crossing angle is larger than the disruption angles,
the beams just travel straight outside the quadrupoles.

The disrupted beams after the IP have an angular spread of about $\pm$
12 mrad ~\cite{TESLATDR,TEL-Snow2005,TELacta2}. The disruption angle
for low-energy particles is proportional to $\sqrt{N/\sigma_z
  E}$~\cite{TEL90,TEL95} and depends very weakly on the transverse
beam sizes. The required crossing angle is determined by the disruption angle, the outer radius of the final quadrupole (about 5
cm~\cite{TEL-Snow2005,TELacta2}), and the distance between the first
quad and the IP (about 4 m), which gives $\alpha_c = 12 +5/400 \approx
25$ mrad.

In the present ILC design~\cite{RDR} only one IP is planned, with a crossing angle of 14 mrad and two detectors in the pull-push configuration.
On the other hand, at the photon collider the crossing angle should be at least 25 mrad. At first sight, it would therefore seem quite reasonable to design the ILC with 25 mrad crossing angle both for the \EPEM\ and the photon
collider.  However, it was decided to make different collision angles due to
very different  requirements to the extraction lines and beam
dumps. In the \EPEM\ case, after collision the beams remain quite
monochromatic, and so there is a possibility to measure their properties
(the energy spectrum and polarization).  At the photon collider, the situation is different:
1) the disrupted beams at a photon collider consist of an equal mixture of
  electrons and photons; 2) beams  have a large angular
 spread  and need exit pipes of a large diameter; 3) the photon beam after the Compton scattering is very narrow, it cannot be dumped  directly at any solid or liquid material. There exist an idea and simulations of a beam dump for the photon collider ~\cite{Telnov-lcws04,TEL-Snow2005,TELacta2}.
Conceptually, it is a long tube, the first 100 m of which is vacuum, followed
by a 150 m long gas converter ending in a water-filled beam dump. In
addition, there are fast sweeping magnet for electrons. Due to a large
beam width, no detailed diagnostics are possible except, perhaps, for beam
profile measurements.

So, it makes sense to have different crossing angles, separate extraction lines and beam dumps for \EPEM\ and \GG.  For the transition from \EPEM\
 to \GG, one has to move the detector and about 700 m of the up-stream
 beamline~\cite{TEL-LCWS06-1}.
\subsection{The laser and optics}
The photon collider at ILC(500) requires a laser system with the
following parameters~\cite{TEL-Snow2005,TELacta2}: flash energy $A
\sim 10$ J, $\sigma_t \sim 1.5$ ps, $\lambda \sim 1$ \MKM ($\le 5 E_0$
[\TEV] \MKM\ in a general case), and the ILC pulse structure: 3000
bunches within a 1 ms train and 5 Hz repetition rate for the trains,
the total collision rate being 15 kHz.

In addition to the average repetition rate, the time structure is of
great importance. The average power required of each of the two lasers
for the photon collider at the ILC is 10 J $\times$ 15000 Hz $\sim$
150 kW; however, the power within the 1 msec train is 10 J $\times
3000/0.001 \sim 30$ MW!  The cost of diodes is about ${\cal O }(1\$)
$/W, the pumping efficiency about 25\%, so the cost of just the pumping diodes
would be
% at least
${\cal O }$(\$100M).

Fortunately, at the PLC and other Compton scattering applications the same laser bunch can be used multiple times. The most attractive approach is a stacking optical cavity that is pumped by a laser via a semi-transparent mirror~\cite{Tfrei,e-e-99,TEL2001,Will2001,TESLATDR,Klemz2005}.  The
ILC pulse structure (3000 bunches in the train with inter-pulse
distance ~100 m) is sufficient to create a ring cavity around the detector. One can create inside such a cavity a light pulse with an intensity that is by a factor of
$Q$ (the quality factor of the cavity) greater than the incoming laser
power. The value of $Q$ achievable at such powers is several hundreds
and (even $Q>1000$ is not excluded). This means reduction of the
required laser power by the large factor $Q$.

The external optical cavity (pulse stacking cavity) idea has proven to
be a highly useful technique for HEP and other application (see the
last Section).  Recently at LAL, F.~Zomer's group has received in a Fabry-Perot cavity an enhancement factor of 10000!~\cite{Zomer-1}  They found that a simple concentric Fabry-Perot cavity is very unstable and sensitive to displacements. Much more stable is the 2D (planar) concave 4-mirror system, but it also has a problem: astigmatic and only linearly (or elliptically) polarized eigenmodes due to different reflection for the s and p waves. A possible solation: 3D 4-mirror cavity that has reduced astigmatism and stable circularly polarized eigenmodes. The LAL group working in collaboration with Japanese colleagues (see T.~Takahashi's talk in these proceedings) is in the process of developing such a 3D 4-mirror cavity, plans to install it at the KEK ATF2 facility and obtain 1 MW average power in the cavity. For the photon collider, very stable both circular and linear polarizations are needed. We see that this is not a simple task.

Recently, J.~Gronberg and B.~Stuart from LLNL have proposed a plan on possible stages in the development and construction of the laser
system for the PLC~\cite{Gronberg-TILC09}. They demonstrate that
all the necessary technologies already exist. Pulse injection and intermediate amplification
devices are off-the-shelf technologies; the main amplifier is not
commercially available but at LLNL all required technologies exist
(the Mercury laser is an existence proof). Gronberg and Stuart specified six stages,
where the first one is the pre-conceptual design and the last one is the
construction of a full-scale cavity and demonstration of its
operation. The rough estimate of the cost of the laser system is \$20
M (``once it is known technology''). These are very nice plans! (dreams?) Unfortunately,
at present there are no resources for such a program, partially due to
the very uncertain plans on the ILC.

In summary: at present, practically all laser technologies and components
required for a photon collider are in existence; nevertheless, the
construction of such a state-of-the-art laser system will not be an
easy task. The next step will be the development (on paper) of a detailed
laser scheme, its optimization, analyses of tolerances, methods of
stabilization, figuring out what already exists and is known and what
should be experimentally verified. Current development of passive stacking
cavities for various applications based on Compton scattering is
very helpful for the PLC.

\section{Photon collider Higgs factory as a precursor to ILC}

 We have routinely assumed that a linear collider would start with \EPEM,
while \GG,\GE\ collisions would arrive several years later. However, on many occasions many people have suggested to build the photon
collider before \EPEM\ (or even without \EPEM) because it is
simpler (no $e^+$, may be no damping rings) and a somewhat lower beam
energy is needed to produce an intermediate-mass Higgs boson.
Several such suggestions (not all) are 1) V.~Balakin et al.
(1993)\cite{Bal-Gin93}, based on the VLEPP and D.~Asner et al.
(2001)~\cite{Asner}, based on CLIC-1 with $E_0=70$ GeV.

Recently, in October 2008, H.~Sugawara gave a talk at the ILCSC meeting entitled
``Photon-photon collider Higgs factory as a precursor to ILC''. His
main motivation was the following: the ILC(500) is too expensive, so let
us build first a collider for the smallest reasonable energy. If the physics goal
is H(120), then it could be produced in $\GG\ \to H$ at $2E_0 \sim 160$ GeV,  while $\EPEM\ \to ZH$ needs higher energy. Besides, \GG\ does not need positrons.
Before Sugawara's presentation at the ILCSC meeting, one of the ILCSC members
sent me the slides of the proposal and asked for an opinion on this
subject. My reply to the ILCSC and to Sugawara-san was the following
(shortly): \\[-6mm]
\bi
\item the cost of such a PLC will be not much cheaper because it needs
  damping rings with small emittance and polarized electrons.
  Polarization is absolutely necessary for the Higgs study.
  A polarized electron gun with small emittance does not exist, therefore damping rings are unavoidable; \\[-6mm]
\item  the laser system for the PLC is not simple, its developments have not
  started yet;\\[-6mm]
\item the H(120) can be be studied much better in $\EPEM\ \to ZH$ at
  $2E_0=230$ GeV;\\[-6mm]
\item the ILC community will never agree with a proposal that shifts the
  start of \EPEM\ experiments by about 5 years;\\[-6mm]
\item the PLC gives a unique possibility to study new physics at the
  LC in two additional modes at a small additional cost,
  gives access to higher masses, but it would be better to plan the
  PLC as the second stage of the ILC (as it was usually assumed).\\[-6mm]
  \ei
  After the meeting, the ILCSC requested the LOI Physics pannel to
  look in more detail into the physics case, and the
  GDE to look at machine designs for this kind of a staged approach to ILC
  construction and operation.  In January 2009, T.~Barklow, J.~Gronberg,
  M.~Peskin and A.~Seryi (BGPS) prepared a draft of the report, which
  was discussed at the expanded Physics panel with invited PLC experts and was not supported.

  In February 2009, the BGPS report was reviewed at an ILCSC meeting and the conclusion was the following:
    ``A 180 GeV gamma-gamma precursor would cost about half that of the
  500 GeV ILC, but would produce much less physics. A better
  alternative for early Higgs studies would be a $\sim 230$ GeV
  \EPEM\ collider for studying the Higgs through ZH production; this
  would be about 30\% more costly than the \GG\ collider. ILCSC
  decided not to pursue the gamma-gamma collider (as the ILC precursor)
  further at this time.''

  The PLC as the first stage of ILC was also discussed  at TILC09 in April 2009: T.~Barklow's plenary  talk~\cite{BarklowTILC09} and the discussion at Joint Physics  session with my introductory talk~\cite{TEL-TILC09-2}. All people have agreed that PLC is certainly necessary, but it would be better to   start with \EPEM(230). Since \EPEM(230) is needed in any case, a PLC precursor results in no cost reduction at all. See also J.~Gronberg's talk at PHOTON-2009~\cite{Gronberg-ph09}.  This is all about the PLC as a precursor to the ILC.

  Nevertheless, proposals to build some type of a "cheap" low-energy PLC continue to resurface.  It could have made sense if the production rates
for some particles were higher than those at \EPEM\ factories.
However, simple estimates show that \EPEM\ B-factories (and especially a future Super B factory) have much higher productivity for any particles of interest.
  So, the PLC should wait for its logical turn at the ILC (or CLIC, or other LC).

\section{Applications based on Compton scattering}

In 1981, when the PLC was proposed, the maximum $e\to\gamma$
conversion efficiency in Compton scattering of a laser light was at
the level $k=N_{\gamma}/N_e \sim 10^{-7}$. For the PLC, $k \sim 1$ was
needed. Our first journal paper was published only from the third
attempt because the editors and reviewers considered the idea to be
unrealistic. After invention of the chirped-pulse technique in 1985~\cite{STRIC},
the required flash energy of several joules and ps duration became a reality. The remaining problem was the repetition rate.
An optical pulse stacking cavity provides a nice solution to this
problem. This technique was known for a long time (the Fabry--Perot resonator)
but its wide application in HEP and other Inverse Compton Scattering (ICS)
applications was triggered by PLC developments.

 Collision of a relativistic electron bunch with an intense
laser pulse generates photon beams with the following characteristics: source is bright  (directional, ultra-fast); scattered light is monochromatic (after collimation); tunable wavelength, like FEL; much less expensive than XFEL; broad energy reach: keV, MeV, GeV TeV;  polarization (useful for $e^+$ generation).

Let us enumerate some of the applications:\\[-6mm]
\bi
\item Medical applications:\\[-6mm]
    \bi
\item Dichromatic imaging: illuminate above and
  below contrast K-edge, digital image subtraction. Established at
  synchrotrons (access limited, expensive \$100s); \\[-6mm]
\item Computer  tomography with monochromatic X-rays.
In mammography conventional X-ray imaging is difficult,
soft-tissue contrast is poor.  Monochromatic X-rays  enable new techniques:
phase contrast imaging 3D with low dose; \\[-6mm]
\ei

\item Fast X-ray materials characterization: composition of materials,
  inspection of trucks and containers; \\[-6mm]
\item Nuclear materials detection (in trucks and containers) by
  monochronatic $\gamma$ beam;\\[-6mm]
\item Defect profiling with $e^+$s. MeV photons produce positrons
  which gather at defects, detection of two 510 keV annihilation
  photons shows the source position (PET-tomography). Directly probe
  material defects.\\[-6mm]
\item Nuclear waste assay. Resonance scattering of 1-5 MeV $\gamma$-quanta
  is a unique fingerprint of nuclides, radioactive and stable nuclides
  can be detected;\\[-6mm]
\item Beam diagnostics and polarimetry at electron accelerators; \\[-6mm]
\item Obtaining polarized $e^+$ for \EPEM\ linear
  colliders~\cite{Omori}. Similar to the undulator source of
  polarized $\gamma$ based on $\sim 150$ GeV main linac beams but needs much
  lower electron energies and is independent of the main collider. There is a very active collaboration named POSIPOL, which has already conducted four workshops on this topic;\\[-6mm]
\item Laser cooling of electrons~\cite{TELlascool} can considerably reduce emittances  of beams and increase the luminosity of photon   colliders, provide a fast beam cooling in
  damping rings~\cite{Ruth} for X-ray production and for $e^+$
  production~\cite{Omori}; \\[-6mm]
\item The Photon Collider.\\[-6mm]
\ei
Already many ICS facilities are under construction in the
world~\cite{Italy}. These activities (all enumerated above and some others) are called by people as the "Compton World Wide Web of Laser Compton." Let us hope that the Photon Collider will be eventually constructed somewhere in the world as well!


\begin{thebibliography}{99}

\bibitem{Monig-photon09} K. M\"{o}nig, talk at Photon09, these proceedings. \\[-5mm]

\bibitem{GKST81} I.F.~Ginzburg, G.L.~Kotkin, V.G.~Serbo, and V.I.~Telnov,
Pizma~ZhETF, {\bf 34} 514 (1981); [JETP~Lett. {\bf 34}, 491 (1982)]. \\[-5mm]

\bibitem{GKST83-84} I.F.~Ginzburg et.al., Nucl. Instrum. Meth. {\bf 205}, 47 (1983); ibid {\bf A219}, 5 (1984). \\[-5mm]

\bibitem{TESLATDR} B.~Badelek et. al., Intern. Journ. Mod. Phys.  {\bf A30}, 5097 (2004), hep-ex/0108012. \\[-5mm]

\bibitem{TELacta1} V.I.~Telnov, Acta Physica Polonica, {\bf B37},
 633 (2006), physics/0602172. \\[-5mm]

\bibitem{TELacta2} V.I.~Telnov, Acta Physica Polonica, {\bf B37},
 1049 (2006); physics/0604108. \\[-5mm]

\bibitem{Tel-Photon07} V.I.Telnov,  Nuclear Physics B (Proc. Suppl.) {\bf 184} 271 (2008). \\[-5mm]


\bibitem{TEL90} V.I.~Telnov, Nucl. Instrum. Meth.,{\bf A294}, 72 (1990). \\[-5mm]

\bibitem{TEL95} V.I.~Telnov, Nucl. Instrum. Meth. {\bf A355}, 3 (1995). \\[-5mm]


\bibitem{RDR} ILC Reference Design Report,ILC-Report-2007-001, arXiv:0712.2361 [physics.acc-ph]. \\[-5mm]


\bibitem{TEL-Snow2005} V.I.~Telnov,  Proc. of 2005 Intern. Linear
   Collider Physics and Detector Workshop and 2nd ILC Acceler.
   Workshop, Snowmass, Colorado, 14-27 Aug 2005, ECONF
   C0508141: PLEN0020,2005,  physics/0512048. \\[-5mm]

\bibitem{Klemz2005}  G.~Klemz, K.~Monig and I.~Will,
 Nucl.\ Instrum.\ Meth.\  {\bf A564}, 212 (2006),  physics/0507078. \\[-5mm]

\bibitem{ChenTel} P.~Chen and V.~I. Telnov, Phys.~Rev.~Lett., {\bf
     63}, 1796 (1989). \\[-5mm]

\bibitem{Tfrei} V.I.~Telnov,
Nucl.~Phys.~Proc.~Suppl. {\bf 82}, 359 (2000), hep-ex/9908005. \\[-5mm]

\bibitem{TEL2001} V.I.~Telnov, Nucl. Instrum. Meth., {\bf A472}, 43
(2001), hep-ex/0010033. \\[-5mm]

\bibitem{TEL-LCWS06-2} V.I.~Telnov, Pramana Journal of Physics, {\bf 69}, 957 (2007), physics/0610285. \\[-5mm]

\bibitem{e-e-99} V.I.~Telnov, Int. J. of Mod. Phys. {\bf A15}, 2577 (2000),
 hep-ex/0003024. \\[-5mm]

 \bibitem{Will2001} I.~Will, T.~Quast, H.~Redlin and W.~Sandner,
  {\em Nucl.\ Instrum.\ Meth.\ } {\bf A472}, 79 (2001). \\[-5mm]


\bibitem{Pak} A.V.~Pak, D.V.~Pavluchenko, S.S.~Petrosyan,
  V.G.~Serbo and V.I.~Telnov,
  Nucl. Phys. Proc. Suppl., {\bf 126}, 379 (2004), hep-ex/0301037. \\[-5mm]

 \bibitem{TEL-mont} V.I.~Telnov, talk at the ECFA workshop on linear
  colliders, Montpellier, France, 12-16 November 2003; slides:
\verb$http://www-h1.desy.de/~maxfield/ggcol/montpellier_talks$ \\
\verb$/Valery_lumispec_MONT1.PDF$  \\[-5mm]

\bibitem{TEL-ggLHC} V.~I.~Telnov, Nucl. Phys. Proc. Suppl.{\bf 179-180} (2008) 81. \\[-5mm]

\bibitem{Makarenko}  V.~Makarenko, K.~Monig and T.~Shishkina,
  Eur.\ Phys.\ J.\  {\bf C 32}, (2003) SUPPL1143.  \\[-5mm]

\bibitem{TEL-TILC09-1} V.I.Telnov, Calibration of energies at the photon collider, talk at TILC09, April 17-21, 2009, Tsukuba, Japan. \verb$http://tilc09.kek.jp$ \\[-5mm]

\bibitem{Telnov-lcws04} L.I.~Shekhtman and V.I.~Telnov, Proc. of
 Intern. Conf. on Linear Colliders (LCWS 04), Paris, France, 19-24 Apr
 2004, {\it Paris 2004, Linear colliders}, v1, p.507; physics/0411253. \\[-5mm]

 \bibitem{Palmer} R.~B. Palmer,
 In {\em DPF Summer Study Snowmass '88: High Energy Physics in the
  1990's, Snowmass, Colo., Jun 27 - Jul 15, 1988\/}, SLAC-PUB 4707.

\bibitem{TEL-LCWS06-1} V.I.~Telnov, Proc. LCWS06, India, March
2006. Pramana Journal of Physics, {\bf 69}, 1177 (2007),
physics/0610287. \\[-5mm]

\bibitem{Zomer-1} F.~Zomer, A. Jeremie, EUROTeV-Report-2008-096-1.

\bibitem{Gronberg-TILC09} J.Gronberg and B.Stuart, Photon Collider Laser Work at LLNL, talk at TILC09, April 17-21, 2009, Tsukuba, Japan. \verb$http://tilc09.kek.jp$ \\[-5mm]

\bibitem{Bal-Gin93} V.~E.~Balakin and I.~F.~Ginzburg, Proc. 2nd
   Intern. Workshop on Physics and Experiments with Linear \EPEM\
   Colliders, Waikoloa, Hawaii, 26-30 Apr, 1993, v2, p. 605. \\[-5mm]

\bibitem{Asner}  D.~Asner {\it et al.}, Eur.\ Phys.\ J.\ C {\bf 28} (2003) 27, hep-ex/0111056. \\[-5mm]

\bibitem{BarklowTILC09} T. Barklow, Gamma-gamma collider physics report, talk at TILC09, April 17-21, 2009, Tsukuba, Japan, \verb$http://tilc09.kek.jp$ \\[-5mm]

\bibitem{TEL-TILC09-2} V.I.Telnov, Introduction to the discussion on Physics case of the PLC as the first stage of ILC, talk at TILC09, April 17-21, 2009, Tsukuba, Japan, \verb$http://tilc09.kek.jp$ \\[-5mm]

\bibitem{Gronberg-ph09} J.~Gronberg, Costs versus benefit of an early photon collider project, these proceedings. \\[-5mm]

\bibitem{STRIC} D.~Strickland and G.~Mourou, {\em Opt.~Commun.}, {\bf 56},
  219 (1985). \\[-5mm]

\bibitem{Omori} S.~Araki {\it et al.}, Conceptual design of a polarised positron source based on laser Compton scattering, arXiv:physics/0509016. F.~Zimmermann {\it et al.}, CLIC polarized positron source based on laser Compton scattering,  CERN-CLIC-NOTE-674; M.~Kuriki {\it et al.}, ILC positron source based on laser Compton,  AIP Conf.\ Proc.\  {\bf 980} (2008) 92. \\[-5mm]

\bibitem{TELlascool} V.I.~Telnov, Phys.~Rev.~Lett.,{\bf 78},  4757 (1997),
 Erratum: Phys.~Rev.~Lett. {\bf 80}, 2747 (1998); V.I.~Telnov,  Nucl.~Instrum.~Meth., {\bf A455},  63 (2000), hep-ex/0001029. \\[-5mm]

 \bibitem{Ruth}  Z.~Huang and R.~D.~Ruth, Phys.\ Rev.\ Lett.\  {\bf 80}, 976 (1998). \\[-5mm]

 \bibitem{Italy} Workshop on Compton sources for X/gamma rays: Physics and Applications, 7-12 Sept, 2008, Alghero, Italy,
    \verb$http://agenda.infn.it/conferenceTimeTable.py?confId=367$ \\[-5mm]

\end{thebibliography}
\end{document}